\documentclass[aps,prd,showpacs,twocolumn,floatfix,superscriptaddress,preprintnumbers]{revtex4}
\usepackage{amsmath}
\usepackage{amssymb}
\usepackage{epsfig}
\usepackage{graphicx}
\usepackage{stmaryrd}
\def\10{$SO(10)$}
\def\21{SU(2) $\otimes$ U(1) }

\def\422{$SU(4) \otimes SU(2) \otimes SU(2)$}
\def\321{SU(3) $\otimes$ SU(2) $\otimes$ U(1)}
\def\lsim{\raise0.3ex\hbox{$\;<$\kern-0.75em\raise-1.1ex\hbox{$\sim\;$}}}
\def\gsim{\raise0.3ex\hbox{$\;>$\kern-0.75em\raise-1.1ex\hbox{$\sim\;$}}}

 \newcommand{\ba}{\begin{array}}
\newcommand{\ea}{\end{array}}
\relax

\def\321{$SU(3)\times SU(2)\times U(1)$}

\begin{document}
\renewcommand{\Huge}{\Large}
\renewcommand{\LARGE}{\Large}
\renewcommand{\Large}{\large}

\title{Flow of hypermagnetic helicity in the embryo of a new phase in the electroweak phase transition} 
\author{P.M. Akhmet'ev} \email{pmakhmet@izmiran.ru} 
\affiliation{Pushkov Institute of Terrestrial Magnetism,\\
  Ionosphere and Radiowave Propagation of the Russian Academy of
  Sciences, IZMIRAN, Troitsk, Moscow region, 142190, Russia}
\author{V.B. Semikoz} \email{semikoz@yandex.ru}
\affiliation{Pushkov Institute of Terrestrial Magnetism,\\
  Ionosphere and Radiowave Propagation of the Russian Academy of
  Sciences, IZMIRAN, Troitsk, Moscow region, 142190, Russia}
\author{D.D. Sokoloff}\email{sokoloff@dds.srcc.msu.su}
\affiliation{Department of Physics, Moscow State University, 119999, Moscow, Russia}

\date{\today}
\begin{abstract}

  The dynamics of the magnetic helicity during the electroweak phase
transition in the early Universe is studied. It is shown that the
boundary surface between symmetric (hypermagnetic) phase and
Maxwellian phase with a broken symmetry is a membrana for the
separation of the magnetic helicity. Assuming the total linking
number of knots of hypermagnetic field is negative, it is proved
that the helicity  rising in the Maxwellian phase is left-handed.

\end{abstract}


\maketitle

The primordial magnetic fields in the early Universe could be 
sufficiently strong not influencing neither expansion of universe 
nor Big Bang nucleosynthesis. If such fields survive under certain
conditions after recombination ($ z <1100 $) they could be as  seed fields
in the galactic dynamo mechanism \cite{review, rrew}.

An essential topological characteristic of the magnetic field
which is global invariant in expanding universe with the
changing structure of magnetic field at different scales is the
magnetic helicity, for example, $ H = \int d ^ 3x (\bf B \cdot
\bf A) $ in the case of the Maxwellian field. By a modern knowledge
\cite{bkandu}, it can significantly affect the evolution of
magnetic fields in galaxies.

The cosmological magnetic field and its helicity can be formed as a
result of phase transitions in the early Universe and, in
particular, in the electroweak phase transition. In this phase
transition the hypermagnetic field converts 
into the Maxwellian  electromagnetic field.

In this paper we try to study how the helicity of hypermagnetic
fields is related to the magnetic helicity of Maxwellian fields during
electroweak phase transition. We show that during this
phase transition on the surface separating the phases a
separation of magnetic helicity exists. The magnetic helicity being
collected in Maxwellian phase in the course of this separation 
is conserved then in  further expansion of the Universe and the subsequent
formation of galaxies. As shown in \cite{SS1, SV}, the phenomenon
is due to neutrino asymmetry and parity violation
($P$-non-invariance) in weak interactions. In the absence of
neutrino asymmetry in an ideal plasma the helicity is conserved.

Let us consider a bubble (an embryo of the Maxwell phase) of the
radius $R$, inside of the hot plasma in the early Universe at the
moment of the electroweak phase transition with the temperature
$T_{EW}\sim 100~{\rm GeV}$. Let us assume that this bubble is
growing with the constant velocity, $R(t)=v(t-t_{EW})$, where the
velocity $v$ itself ($v=0.1-1$ according to \cite{Kibble}) is
unessential (cancelled) in the solution to our problem. 

It is important for our calculation that the value $(t-t_{EW})/t_{EW}\ll
1$ is small, or that the temperature during the phase
transition remains constant at the moment
$t_{EW}=M_0/2T_{EW}^2=0.23\times10^{-10}~c$ , where $M_0=M_{Pl}/1.66\sqrt{g^*}$ is
given by the  Plank mass $M_{Pl}=1.2\times 10^{19}~GeV$ and
by the degree of freedom $g^*\sim 100$. This implies that the radius
of the bubble is much less than the horizon size
($2t_{EW}=l_H=1.44~cm$), $R\ll l_H$. More precisely, we shall
assume that the radius of the bubble is much less than the scale of
the mean hypermagnetic field, $R\ll \eta_Y/\alpha_Y\ll l_H$.

In the rest frame of the medium as a whole
the induction equation governing  hypermagnetic fields ${\bf
B}_Y=\nabla\times {\bf Y}$ outside of the bubble reads:
\begin{equation}\label{Faraday} \frac{\partial {\bf B}_Y}{\partial
t}=\nabla \times \alpha_Y {\bf B}_Y + \eta_Y \nabla^2{\bf B}_Y,
\end{equation}
while the evolution equation for Maxwellian field
 ${\bf B}=\nabla\times
{\bf A}$ within the bubble is analogous, but with different
value of the parameter $\alpha$. This parameter characterizes
$P$-non-invariance of weak interactions and it is a
scalar, but not a pseudoscalar like the corresponding parameter
$<{\bf v}\cdot(\nabla\times {\bf v})>$ in the standard
magnetohydrodynamics \cite{bkandu}.

After the phase transition such a parameter of the helicity
is the following \cite{SS1}:
\begin{equation}\label{alpha1}
\alpha =2 \times 10^{-2}G_F T \lambda^{-1}
 \sum_ac_a^{(A)}\xi_{\nu_a},
\end{equation}
while before the  phase transition it takes the form
\cite{SV}:
\begin{equation}\label{alpha}
 \alpha_Y
=3\times 10^{-3} g^{{'}2}\sigma^{-1}T\sum_a\xi_{\nu_a},
\end{equation}
where $G_F$ is the Fermi constant, $T$ is the temperature,
$\lambda$ is the spatial parameter of the inhomogeneity of the
neutrino gas; $c_a^{(A)}=\mp 0.5$ is the axial constant of weak
interactions (the upper sign for electron neutrino); $g^{'}$ is
the coupling constant for hypercharge field in the Weinberg-Salam
model; $\xi_{\nu_a}=\mu_{\nu_a}/T$ is the dimensionless
chemical potential of neutrino, $a=e,\mu,\tau$, and coefficients of
hypermagnetic diffusion
 $\eta=\eta_Y =(4\pi \sigma)^{-1}$ are given by the conductivity of plasma $\sigma \sim 100~T$
and, practically, coincide for both phases. All coefficients
 $\xi_{\nu_a}$, $\alpha_Y$, $\eta_Y$ depend on the temperature (or time) by 
the Friedman law, but for given problem 
at the fixed temperature of the phase transition
$T_{EW}$, these parameters remain unaltered.

Multiplying Eq. (\ref{Faraday}) and its analogue for Maxwellian field 
by the corresponding vector potential and adding the
analogous construction produced by evolution equation 
governing the vector potential (multiplied by hypermagnetic
or magnetic field) after the integration over
the space we get the evolution equation for  the total
helicity
 $H=\int({\bf
B}\cdot{\bf A})d^3x+\int ({\bf B}_Y\cdot {\bf Y})d^3x$, where the
integration is carried over the domains with Maxwell phase and
hypermagnetic field correspondingly. This equation takes the form:
\begin{eqnarray}\label{evolution}
\frac{{\rm d}H}{{\rm dt}}=&&
-2\int ({\bf E}\cdot{\bf B}){\rm d}^3x- \nonumber\\&&
 -\oint_S(({\bf E}\times {\bf A} + A_0{\bf B})\cdot {\bf
n}){\rm d}^2S  +\dots,
\end{eqnarray}
where the dots mean analogous terms for hypermagnetic
field. We take into account the surface integrals which are
omitted in problems for a monophase medium
 \cite{Priest} as integrals over an infinite boundary of the domain.
In our problem namely these integrals determine the flow of the
helicity through the boundary of a bubble of the radius $R$, on
which a separation of the helicity takes place. Accounting for
the boundary condition $A_{\mu}=\cos \theta_{W}Y_{\mu}$, where
$\sin^2\theta_W=0.23$ is the parameter of the standard
Weinberg-Salam model, the integrals above are calculated over the
surface as following:
\begin{equation}\label{sum}
\frac{{\rm d}H_Y}{{\rm dt}}=-\sin^2\theta_W\oint({\bf E}_Y\times {\bf
Y} + Y_0{\bf B}_Y){\bf n}_Y{\rm d}^2S,
\end{equation}
where the unit normal vector ${\bf n}_Y=-\hat{{\bf
e}}_r=(-1,0,0)$ is directed inwards the bubble with Maxwellian phase (the
phase with broken symmetry).

The flow of hypermagnetic helicity density, penetrated
inside the bubble through the surface at the moment of electroweak
phase transition, is the pseudovector given by the formula
\begin{equation}\label{flux}
{\bf S}={\bf n}_Yh_Y(t)={\bf n}_Y\left(\frac{1}{4\pi
R^2(t)d}\right)\int_{t_{EW}}^t{\rm dt}\frac{{\rm d}H_Y(t)}{{\rm
dt}}.
\end{equation}
This flow is analogous to the vector flow of the energy of a flat electromagnetic
wave
 ${\bf S}=W{\bf n}$, where $W=(E^2 +
B^2)/8\pi$ is the energy density of the field. Here  
$4\pi dR^2(t)$ is the volume of a thin spherical layer
with the thickness $d$ of the domain wall separating the two
phases. For fields at the scale $R$,
 $Y\sim
B_YR$, the value of the flow is inversely to the thickness $d$ ,
$h_Y\sim d^{-1}$. 

It is not difficult to prove that the surface
integrals are equal to zero, i.e. there is no separation of the
helicity if we substitute the flat hypermagnetic field, $Y_0=Y_z=0$,
$Y_x=Y(t)\sin k_0z$, $Y_y=Y(t)\cos k_0z$, (see estimates of the
baryon asymmetry in  \cite{SSV}). But in the case of 3D-
field with nonzero helicity the considered integrals are
nontrivial. Let us consider the following potential of the
hypermagnetic field with the number of the linked loops equal to
$n$ :
\begin{eqnarray}\label{3D}
Y_r(t,\rho,\theta)=&&\frac{-Y(t)\cos\theta}{(\rho^2 +1)^2},\nonumber\\
Y_{\theta}(t,\rho,\theta)=&&\frac{Y(t)\sin\theta }{(\rho^2 +1)^2}\Bigl[1 + B(\rho-1)^2 +
b(\rho-1)^3\Bigr],\nonumber\\
Y_{\phi}(t,\rho,\theta)=&&\frac{-Y(t)n\sin\theta}{(\rho^2 +1)^2}\Bigl[\rho + C(\rho-1)^2
+ \nonumber\\&& + (C+c)(\rho-1)^3\Bigr],
\end{eqnarray}
where $\rho=r/R$, dependence on  time is $Y(t)=2B_0(t)/\pi R$
(compare \cite{Giovannini2}), and the   coefficients $b,c, B, C$ are
calculated below. We correct here the misprint in our paper \cite{ASS} where in the same Eq. (\ref{3D}) we missed the amplitude $Y(t)$ for the component $Y_{\theta}(t,\rho,\theta)$.

The hypermagnetic field ${\bf B}_Y=\nabla\times {\bf Y}$ near the surface of phase separation, 
$0<\rho -1\ll 1$, has the components
\begin{eqnarray}\label{hypermag3D}
&&B^Y_r=\frac{1}{r\sin \theta}\left[\frac{\partial}{\partial \theta}(\sin \theta)Y_{\phi}\right]=-\frac{Y(t)n\cos\theta}{2R(t)}\times\nonumber\\&&\times
\Bigl[1 - 2(\rho -1) + (C+2)(\rho -1)^2 + O((\rho -1)^3)\Bigr],\nonumber\\&&
B_{\theta}^Y=-\frac{1}{r}\left[\frac{\partial}{\partial r}(rY_{\phi})\right]=\frac{Y(t)n\sin\theta}{2R(t)}\Bigl[(\rho-1)(C-1)\nonumber\\&&+
(\rho -1)^2\left(\frac{3}{2}c +\frac{5}{2} - C\right)+O((\rho -1)^3)\Bigr],\nonumber\\&&
B_{\phi}^Y=\frac{1}{r}\left[\frac{\partial}{\partial r}(rY_{\theta}) - \frac{\partial Y_r}{\partial \theta}\right]=\frac{Y(t)\sin\theta}{4R(t)}\Bigl[-2 + \nonumber\\&& +(\rho -1)(4+2B)+(\rho -1)^2(3b-3 - 5B)+\nonumber\\&&+O((\rho -1)^3)\Bigr]~.
\end{eqnarray}
At the surface of bubble $\rho=1$ our
potential (\ref{3D}) and corresponding hypermagnetic field (\ref{hypermag3D}) are like in
paper \cite{Giovannini2}. Obviously, $\nabla\cdot {\bf
B}_Y=0$. We used the Lorentz gauge $\partial Y_{\mu}/\partial
x_{\mu}=0$ to calculate the temporal component of hypercharge field

$$
Y_0(\rho,\theta, t)=-\frac{4\rho\cos\theta}{(\rho^2 +1)^3}\int_{t_{EW}}^t
\frac{Y(t^{'})}{R(t^{'})}{\rm d}t^{'}.
$$

A straightforward calculation of the surface term (\ref{sum})
gives the following equation:
\begin{equation}\label{sum2}
\frac{{\rm d}H_Y(t)}{{\rm dt}}=\frac{2\pi\sin^2\theta_W n}{3}R(t)Y(t)\int_{t_{EW}}^t
\frac{Y(t^{'})}{R(t^{'})}{\rm d}t^{'},
\end{equation}
where we substituted in the expression ${\bf E}_Y=-\partial {\bf Y}/\partial t - \nabla Y_0$ the gradient $\nabla Y_0$,
\begin{eqnarray}
\nabla Y_0=&&\frac{1}{R(t)}\int_{t_{EW}}^t {{Y(t^{'}){\rm
d}t^{'}} \over {R(t^{'})}}\Bigl[\frac{4\sin\theta \hat{{\bf e}}_{\theta}}{(\rho^2
+1)^3}-\nonumber\\&&-\frac{4\cos \theta (1-5\rho^2)\hat{{\bf e}}_r}{(\rho^2+1)^4}\Bigr],
\end{eqnarray}
and took into account that in the case of the axial-symmetric
configuration (\ref{3D}) the vector $B_r^Y$ is independent of the
coordinate $\phi$. The values in Eq. (\ref{sum2})  including $\nabla Y_0\times {\bf Y}=\hat{{\bf e}}_r(\nabla Y_0)_{\theta}Y_{\phi}$ are calculated at
the surface of bubble $\rho=1$.

Hence the problem is reduced to the calculation
 $Y(t)$ from the Faraday equation (\ref{Faraday}) which for the considered potential (\ref{3D}) and hypermagnetic field
 (\ref{hypermag3D}) at the boundary $\rho=1$ can be rewritten
by components as
\begin{eqnarray}\label{Faraday3}
&&\frac{\partial B_r^Y}{\partial t} -\eta_Y(\nabla^2 {\bf B}_Y)_r=\frac{\alpha_Y}{r\sin\theta}\frac{\partial}{\partial \theta}(\sin\theta B_{\phi}^Y)=\nonumber\\&&=-\frac{\alpha_Y Y(t)\cos\theta}{R^2(t)}, \nonumber\\&&\frac{\partial B_{\theta}^Y}{\partial t} -\eta_Y(\nabla^2 {\bf B}_Y)_{\theta}=-\frac{\alpha_Y}{r}\frac{\partial}{\partial r}(rB_{\phi}^Y)=\nonumber\\&&=-\frac{\alpha_Y Y(t)\sin\theta}{2R^2(t)}(B+1)=0,\nonumber\\&&\frac{\partial B_{\phi}^Y}{\partial t} -\eta_Y(\nabla^2 {\bf B}_Y)_{\phi}=\frac{\alpha_Y}{r}\left[\frac{\partial}{\partial r}(rB_{\theta}^Y) -\frac{\partial B_r^Y}{\partial \theta}\right]=\nonumber\\&&=\frac{\alpha_YY(t)n\sin\theta}{2R^2(t)}(C-2)~.
\end{eqnarray}
Let us pay attention to the zeroth result for $B_{\theta}^Y$ at the boundary $\rho=1$ (see also in Eq. (\ref{hypermag3D}))
that forces us to choose $B=-1$ in Eq. (\ref{Faraday3})) for that component. The Laplacian in the l.h.s. of Faraday equation for the same 
component also vanishes at the boundary $\rho=1$, $(\nabla^2{\bf B}_Y)_{\theta}=(5 +3c)Y(t)n\sin\theta/2R^3(t)=0$,
if we choose $c=-5/3$.

Then accounting for the other Laplacian components on the same boundary surface, the radial one, $(\nabla^2{\bf B}_Y)_r=Y(t)n\cos\theta (2-C)/R^3(t)\neq 0$, and the zeroth (under conditions $b=-2, B=-1$) azimuthal component,
$(\nabla^2{\bf B}_Y)_{\phi}=Y(t)\sin\theta(3b-3B+3)/2R^3(t)=0$, one gets from the first equation (\ref{Faraday3}) for radial component,
\begin{equation}\label{radial}
\frac{\dot{Y}}{Y} - \frac{\dot{R}}{R}=\frac{2\alpha_Y}{nR} - \frac{\eta_Y(4-2C)}{R^2},
\end{equation}
while from the third equation (\ref{Faraday3}) for the azimuthal component we find
\begin{equation}\label{azimuthal}
\frac{\dot{Y}}{Y} - \frac{\dot{R}}{R}=\frac{\alpha_Yn(2-C)}{R}~.
\end{equation}
Thus, in addition to the parameters $B=-1$,
$c=-5/3$, $b=-2$  combining eqs. (\ref{radial}) and (\ref{azimuthal}) we find the last parameter
$C(t)$ in Eq. (\ref{3D}),
\begin{equation}\label{C}
C(t)=\frac{2(n- n^{-1})\alpha_Y R^{-1} + 4\eta_Y R^{-2}}{n\alpha_Y
R^{-1} + 2\eta_Y R^{-2}}.
\end{equation}

Substituting the parameter (\ref{C}), e.g., into Eq. (\ref{azimuthal}) one obtains the ordinary differential equation for
the amplitude $Y(t)$,
\begin{equation}\label{Y1}
\frac{\dot{Y}(t)}{Y(t)}-\frac{\dot{R}(t)}{R(t)}=\frac{ 2\alpha^2_Y}{n\alpha_Y R(t) + 2\eta_Y}.
\end{equation}

In the realistic situation of finite conductivity a scale of the
mean hypermagnetic field $\Lambda=\kappa \eta_Y/\alpha_Y$, where
$\kappa\geq 1$, should be much bigger than the diameter of the
bubble in the new phase, i.e. the following inequality has to be
satisfied:
 $\alpha_Y  R(t) \ll \kappa \eta_Y $. If a more stronger condition $\alpha_Y  R(t)\ll 2\eta_Y/n\leq \kappa
\eta_Y $ is fulfilled, then from (\ref{Y1}) for the function
$B_Y(t)=Y(t)/R(t)$  using (\ref{alpha}) we get
\begin{eqnarray}\label{dynamo}
&&B_Y(t)=B_Y(t_{EW})\exp
\left[\left(\frac{\alpha^2_Y}{\eta_Y}\right)(t-t_{EW})\right]=\nonumber\\
&&=B_Y(t_{EW})\exp\left[63\left(\frac{\xi_{\nu}}{0.001}\right)^2\frac{(t-t_{EW})}{t_{EW}}\right],
\end{eqnarray}
where  $B_Y(t_{EW})$ is the hypermagnetic field amplitude  
on the scale of the bubble, $\alpha_Y=\alpha_Y(T_{EW})$,
$\eta_Y=\eta_Y(T_{EW})$ are the constant coefficients at the
moment of the phase transition, the sum
$\xi_{\nu}=\sum_a\xi_{\nu_a}(T_{EW})$ is the net neutrino asymmetry
(neutrino degeneracy parameter); $(t-t_{EW})/t_{RW}\ll 1$ is a small parameter for self-consistency
of our problem (see above).

Substituting the amplitude of the hypercharge field
$Y(t)=B_Y(t)R(t)$ on the surface of the phase separation
 (\ref{dynamo}) into the expression of the surface integral   (\ref{sum2}), after the integration over time
and division by the volume  of the spherical layer with the
thickness $d$ we get from
 (\ref{flux}) the value of the flow of hypermagnetic helicity density
 through the surface of the bubble,
\begin{equation}\label{surface} \frac{h_Y(t)}{G^2cm}= \frac{5\times 10^{-3} n}{ d(cm)}\left(\frac{B_Y(t_{EW})}{1~G}\right)^2\left(\frac{t-t_{EW}}{t_{EW}}\right)^2~.
\end{equation}
An unknown neutrino asymmetry at the moment of the phase
transition is estimated by
 $(\xi_{\nu}/0.001)\simeq 0.12$.
This estimate corresponds to the restrictions in
 Eq. (24) in paper \cite{SV}, obtained from the condition
that the hypermagnetic field survives against
ohmic diffusion for spatial scales $\sim\eta_Y/\alpha_Y$.

Let us note that in order to avoid the screening of the
hyperelectric field ${\bf E}_Y$ and the temporal component  $Y_0$
over the surface of the bubble, the thickness $d$ of the domain
wall should be less than the Debye radius, $d< r_D=\sqrt{3T_{EW}/4\pi e^2
n_e}\sim 10/T_{EW}$, that allows to estimate  the
factor $d^{-1}$ in the formula (\ref{surface}) as
$d^{-1}(cm)>10^{15}/2$. This means that a moderate hypermagnetic
field $B_Y(t_{EW})$ provides a huge flow of the helicity density (\ref{surface}).

Indeed, substituting into
 (\ref{surface}) the value of hypermagnetic field at the moment of phase transition $B_Y(t_{EW})$
  estimated in \cite{SSV} as $B_Y(t_{EW})\sim 5\times 10^{17}~G$, one gets  $h/G^2cm>
6.25\times 10^{47}[(t-t_{EW})/t_{EW}]^2$. Such huge value estimated at the moment
of the growth of a bubble of the new phase, e.g, for
$R(t)/l_H<[(t-t_{EW})/t_{EW}]\sim 10^{-6} $, accounting for 
the following conservation of the net global helicity  summed over
different protogalactic scales, occurs much bigger than the helicity density
of galactic magnetic field  $h_{gal}\sim 10^{11}~G^2cm$, (see
also estimates of the primordial magnetic helicity in paper
\cite{SS2}).

We have to note that a growth of hypermagnetic field before the
electroweak phase transition depends essentially (exponentially) 
on the neutrino asymmetry
 ($B_Y(t)= B_0^Y\exp [\int_{t_0}^t(\alpha_Y^2(t^{'})/4\eta_Y(t^{'}))dt^{'}]$ in $\alpha^2$-dynamo \cite{SV}).
 But in the expression for helicity (\ref{surface}) a hyper-magnetic field $B_Y(t)\approx B_Y(t_{EW})$ is fixed
at the moment of the phase transition, moreover, for a small bubble
the answer is practically independent of neutrino asymmetry in the
time-depended field $B_Y(t)$ given by (\ref{dynamo}).

The single bubble of the Maxwellian phase inside of ambient symmetric
phase with the potential given by Eq. (\ref{3D}) near
the boundary, is a reasonable approximation during the beginning
of the phase transition before percolation (junction of bubbles).
One can consider also another final step of the phase transition,
when a new phase with broken symmetry prevails and a single bubble of
the symmetric phase with hypermagnetic field inside exists. It is not
hard to check that in this case the change of sign $\rho -1>0$ to
$\rho-1<0$ in the potential (\ref{3D}) gives the same components
of hypermagnetic field inside the bubble $\rho<1$. Let us note that
in the considered approximation (\ref{3D}) magnetic charges near the
surface of the phase transition and over this surface itself are absent, 
$\nabla\cdot {\bf B}_Y=0$.

A choice of the negative sign of the helicity density 
 (\ref{surface}) if $n<0$ corresponds to the result
\cite{Vachaspati} for the left-handed magnetic helicity in the
same electroweak phase transition. That result is obtained for
the mechanism of decay of linked loops of $Z$-strings leading 
to creation of magnetic monopole-antimonopole pairs at the ends of 
a decaying string, after which the reconnection of each such pair proceeds through 
junction by loops of Maxwellian field.

Let us recall that the pseudoscalar $n$ is the number of pairs of
linking magnetic field loops entering the Gauss integral for magnetic helicity, 
 $H(t)=\int d^3xh(t, {\bf x})=n\Phi_1\Phi_2$ \cite{Priest}.
This pseudoscalar changes the sign after one of the loops in a pair
changes the sign (direction)  of the flow ${\bf \Phi}_i$.

For a single bubble of the symmetric phase the flow of the helicity density 
 through the surface (\ref{flux}) preserves the value (\ref{surface}). Moreover, 
 for the same $n<0$ this flow 
does not change the negative sign after the direction of the flow is changed, ${\bf n}_Y\to -{\bf
n}_Y=\hat{\bf e}_r=(1,0,0)$. This well corresponds to the
meaning of the problem: magnetic helicity of the Maxwellian field
rises, unless helicity of the hypermagnetic field inside
the bubble goes down.

\end{document}